\documentclass[12pt]{article}
\usepackage{graphicx}
\usepackage{graphics}
\title{Integral representations of functions and Addison-type series for mathematical constants} 
\author{Mark W. Coffey\\
Department of Physics\\
Colorado School of Mines\\
Golden, CO  80401\\
(Received $\mbox{~~~~~~~~~~~~~~~~~~~~~~~~~~~~~~~2010}$)}
\date{May 17, 2010}
\begin{document}
\maketitle
\baselineskip=25 pt
\begin{abstract}

We generalize techniques of Addison to a vastly larger context.
We obtain integral representations in terms of the first periodic Bernoulli
polynomial for a number of important special functions including the Lerch
zeta, polylogarithm, Dirichlet $L$- and Clausen functions.  These results then enable a variety of Addison-type series representations of functions.  Moreover, we
obtain integral and Addison-type series for a variety of mathematical constants.

\end{abstract}
 
\medskip
\baselineskip=15pt
\centerline{\bf Key words and phrases}
\medskip 

\noindent

Lerch zeta function, Hurwitz zeta function, polylogarithm function, 
Dirichlet $L$ functions, Clausen functions, generalized Somos constants, Glaisher-Kinkelin constant, Kinkelin constant, Stieltjes constants, series representation

\vfill
\centerline{\bf 2010 AMS codes} 
11M06, 11Y60, 11M35

\baselineskip=25pt
\pagebreak
\medskip
\centerline{\bf Introduction and statement of results}
\medskip

Very recently we have shown how a method of Addison \cite{addison} for 
series representations of the Euler constant $\gamma$ may be generalized in 
several directions \cite{coffeyjnt}.  In particular, we demonstrated generalizations to the important Stieltjes constants $\gamma_k(a)$ of analytic number theory. 
In this paper, we present integral representations for a variety of very useful special functions.  These representations are such that Addison-type series
for function values and other mathematical constants may be developed and we
illustrate this in several instances.  The presentation demonstrates the wide
applicability of our methods.

We let $\Phi(z,s,a)$ be the Lerch zeta function and Li$_s(z)=z\Phi(z,s,1)$
be the polylogarithm function \cite{sri}.  Integral representations, 
including with contour integrals, are known for $\Phi$, permitting, among other
topics, analytic continuation.  
We let $\zeta(s,a)=\Phi(1,s,a)$ be the Hurwitz zeta function, $\zeta(s)=
\zeta(s,1)$ be the Riemann zeta function \cite{edwards,ivic,riemann,titch}, $\psi=\Gamma'/\Gamma$ be 
the digamma function (e.g., \cite{nbs}), $\psi^{(k)}$ be the polygamma functions \cite{nbs}, and $_pF_q$ the generalized hypergeometric function \cite{andrews}.

We let $P_1(x)=B_1(x-[x])=x-[x]-1/2$ be the first periodized Bernoulli polynomial,
and $\{x\}=x-[x]$ be the fractional part of $x$.  Being periodic, $P_1$ has the
Fourier series \cite{nbs} (p. 805)
$$P_1(x)=-\sum_{j=1}^\infty {{\sin 2\pi j x} \over {\pi j}}.  \eqno(1.1)$$
Addison \cite{addison} gave an interesting series representation for $\gamma$ \cite{gerst},
$$\gamma={1 \over 2}+{1 \over 2}\sum_{n=1}^\infty n \sum_{m=2^{n-1}}^{2^n-1}{1 \over
{2m(m+1)(2m+1)}}=1-{1 \over 2}\sum_{n=1}^\infty n \sum_{m=2^{n-1}+1}^{2^n} {1 \over {m (2m-1)}}.  \eqno(1.2)$$
His approach uses an integral representation for the Riemann zeta function in terms of $P_1$.  Various transformations of Addison's result are possible, including \cite{gerst,hardy,sandham,vacca}
$$\gamma=1+\sum_{j=3}^\infty (-1)^j {{[\ln(j-1)/\ln 2]} \over j}=\sum_{j=1}^\infty
(-1)^j {{[\ln j/\ln 2]} \over j}=1+\sum_{j=1}^\infty {{(-1)^j} \over j}
\left\{{{\ln j} \over {\ln 2}}\right\}.  \eqno(1.3)$$
Therefore, Addison's result can be connected to other topics, including the
binary expansion of $\gamma$.  

In this paper, we present many significant generalizations of Addison's result.  
As with the example (1.3), our results can then be connected with the binary or
$n$-ary expansion of function values and other fundamental mathematical constants.

{\bf Proposition 1}.  We have 
$$\Phi(z,s,a)={1 \over a^s}+{z \over {2(a+1)^s}}+\int_1^\infty {z^x \over {(x+a)^s}
}dx+\int_1^\infty \left[{{z^x \ln z} \over {(x+a)^s}}-{{sz^x} \over {(x+a)^{s+1}}}
\right]P_1(x)dx.  \eqno(1.4)$$
This representation holds for $a \in C/\{-1,-2,\ldots\}$ and $s \in C$ when 
$|z|<1$ and for Re $s>1$ when $|z|=1$.  
From this result follows many others, including for the polylogarithm function:
{\newline \bf Corollary 1}.  We have
$$\mbox{Li}_s(z)={z \over 2}+\int_1^\infty {z^x \over x^s}dx +\int_1^\infty
\left[{z^x \over x^s}\ln z- s{z^x \over x^{s+1}} \right]P_1(x)dx.  \eqno(1.5)$$
Proposition 1 and Corollary 1 are applied in the following to develop integral representations for other special functions and series representations for fundamental mathematical constants.

An immediate consequence of Proposition 1 is a connection with a specific
generalized hypergeometric function $_{k+1}F_k$.  We have for integers $k \geq 0$,
{\newline \bf Corollary 2}
$$\Phi(z,k,a)={1 \over a^k} ~_{k+1}F_k(1,a,\ldots,a;a+1,\ldots,a+1;z)$$
$$={1 \over a^k}+{z \over {2(a+1)^k}}+\int_1^\infty {z^x \over {(x+a)^k}
}dx+\int_1^\infty \left[{{z^x \ln z} \over {(x+a)^k}}-{{kz^x} \over {(x+a)^{k+1}}}
\right]P_1(x)dx.  \eqno(1.6)$$

In turn, Corollary 1 implies other results.  As an example, we have the following
for the dilogarithm and polylogarithm functions.
{\newline \bf Corollary 3}.  Let Re $\alpha>-1$.  Then (a)
$$\int_0^1 t^{\alpha-1}\mbox{Li}_2(t)dt={1 \over \alpha}\zeta(2)-{1 \over 
\alpha^2}[\psi(\alpha+1)+\gamma],  \eqno(1.7a)$$
recovering a known result (e.g., \cite{sri}, p. 109).  
Let $_2F_1$ be the Gauss hypergeometric function, $n\geq 2$ an integer, and
$h(s) \equiv -\zeta(s)/s+(s+1)/[2s(s-1)]$.  Then we have (b)
$$\int_0^1 t^{\alpha-1}\mbox{Li}_n(t)dt={1 \over {2(\alpha+1)}}+{1 \over n}
~_2F_1(1,n;n+1;-\alpha)+(-1)^{n+1} \sum_{j=1}^n {{(-1)^j j} \over \alpha^{n-j+1}}
h(j)$$
$$+{{(-1)^{n+1}} \over \alpha^n} \left[\psi(\alpha+1)-\ln(\alpha+1)+{1 \over {2(\alpha+1)}}\right].  \eqno(1.7b)$$
 
We may note the value
$\zeta(3)$ on the right side of (1.7a) when $\alpha\to 0$, owing to the 
expansion $\psi(\alpha+1)=-\gamma+\zeta(2)\alpha-\zeta(3)\alpha^2+\zeta(4) \alpha^3
+O(\alpha^4)$.  Indeed, the proof of Corollary 3 shows how other general
results for integrals $\int_0^1 t^{\alpha-1}\mbox{Li}_s(t)dt$ may be obtained.

We recall definitions of the generalized Clausen function Cl$_n$ for nonnegative
integers $n$:
$$\mbox{Cl}_n(\theta)=\mbox{Im} ~\mbox{Li}_n(e^{i\theta}), ~~~~\mbox{for}~n~
\mbox{even},$$
$$~~~~~~~~=\mbox{Re} ~\mbox{Li}_n(e^{i\theta}), ~~~~\mbox{for}~n~\mbox{odd}. \eqno(1.8)$$
Therefore, from Corollary 1 we obtain the following.
{\newline \bf Corollary 4}.  We have (a)
\newline{for $n$ even:}
$$\mbox{Cl}_n(\theta)={1 \over 2}\sin \theta+\int_1^\infty {{\sin x\theta} \over
x^n}dx+\int_1^\infty {1 \over x^n}[\theta \cos x\theta-{n \over x}\sin x\theta]
P_1(x)dx,  \eqno(1.9a)$$
\newline{for $n$ odd:}
$$\mbox{Cl}_n(\theta)={1 \over 2}\cos \theta+\int_1^\infty {{\cos x\theta} \over
x^n}dx-\int_1^\infty {1 \over x^n}[\theta \sin x\theta-{n \over x}\cos x\theta]
P_1(x)dx,  \eqno(1.9b)$$
(b)
$$\mbox{Cl}_2(\theta)={3 \over 2}\sin \theta-\theta \mbox{Ci}(\theta)+\int_1^\infty {1 \over x^2}[\theta \cos x\theta-{2 \over x}\sin x\theta]P_1(x)dx,  \eqno(1.10)$$
and (c) for the Catalan constant $G=$Cl$_2(\pi/2)$,
$$G={3 \over 2}-{\pi \over 2}\mbox{Ci}\left({\pi \over 2}\right)+\int_1^\infty {1 \over x^2}\left[{\pi \over 2}\cos \left({{\pi x} \over 2}\right)-{2 \over x}\sin \left({{\pi x} \over 2}\right) \right]P_1(x)dx.  \eqno(1.11)$$
In parts (b) and (c), Ci$(z) \equiv -\int_z^\infty (\cos t)(dt/t)$ is the cosine
integral.  

The representations of parts (a) and (b) permit the recovery of known 
properties, including the duplication formula ${1 \over 2}$Cl$_2(2\theta)=$Cl$_2
(\theta)-$Cl$_2(\pi-\theta)$.  The Clausen functions are useful in mathematical
physics (e.g., \cite{coffeyjmp}).

In \cite{coffeyjnt} we presented series representations for the digamma function.
In particular, since $H_n=\psi(n+1)+\gamma$, $\gamma=-\psi(1)$, for integers
$n \geq 1$, we effectively obtained series representations for the harmonic numbers
$H_n\equiv \sum_{p=1}^n 1/p$.  Therefore, by integration we obtain results
for $\ln \Gamma$.  For instance, we have
{\newline \bf Corollary 5}.  We have for Re $z>0$,
$$\ln \Gamma(z)=\left(z-{1 \over 2}\right)-z+1-{1 \over 4}\sum_{n=0}{1 \over 2^n}
\sum_{j=0}^\infty [\ln(z+bj)-\ln(1+bj)+\ln(z+b+bj)-\ln(1+b+bj)$$
$$-2\ln(2z+b+2bj)+2\ln(2+b+2bj)]_{b=2^{-n}}, \eqno(1.12)$$
that follows from Proposition 3(a) of \cite{coffeyjnt}.  Thereby, we obtain
series for constants such as $\ln \sqrt{\pi}$.  
Additionally, if we were to differentiate, for example, (1.8) in \cite{coffeyjnt},
we obtain series for polygammic constants such as $\psi'(1)=\zeta(2)$.  More
generally, we may find series including for 
$$\psi^{(j)}(1)=(-1)^{j+1}j!\zeta(j+1).  \eqno(1.13)$$
The numbers of (1.13) are cases of the relation $\psi^{(n)}(x)=(-1)^{n+1}n!\zeta(
n+1,x)$, and, as is well known, are closely related to generalized harmonic
numbers $H_n^{(r)} \equiv \sum_{j=1}^n {1 \over j^r}$,$H_n \equiv H_n^{(1)}$:
$$H_n^{(r)}={{(-1)^{r-1}} \over {(r-1)!}}\left[\psi^{(r-1)}(n+1)-\psi^{(r-1)}
(1) \right ].  \eqno(1.14)$$

We have
{\newline \bf Proposition 2}.  Let Re $s>0$, Re $a>0$, and $k\geq 2$ be an integer.
Then we have the series representation  
$$\zeta'(s,a)+{a^{1-s} \over {(s-1)^2}}+{{\ln a} \over {2a^s}} + {a^{1-s} \over
{s-1}}\ln a$$
$$=\sum_{n=0}^\infty {1 \over k^n}\sum_{j=0}^\infty \left\{{1 \over 2}\left({1 \over k}-1\right)\left[{{\ln(bj+a)} \over {(bj+a)^s}} + {{\ln(b(j+1)+a)} \over {(b(j+1)+a)^s}}\right]+{1 \over k}\sum_{m=1}^{k-1} {{\ln[b(j+m/k)+a]} \over {[b(j+m/k)+a]^s}}\right\}_{b=k^{-n}}.  \eqno(1.15)$$ 
This Proposition engenders many Corollaries.  
Taking the limit as $s\to 1$ in (1.15) gives
{\newline \bf Corollary 4}.  We have for Re $a>0$
$$-\gamma_1(a) +{1 \over 2}{{\ln a} \over a}=\sum_{n=0}^\infty {1 \over k^n}\sum_{j=0}^\infty \left\{{1 \over 2}\left({1 \over k}-1\right)\left[{{\ln(bj+a)} \over {(bj+a)}} + {{\ln(b(j+1)+a)} \over {(b(j+1)+a)}}\right]\right.$$
$$\left.+{1 \over k}\sum_{m=1}^{k-1} {{\ln[b(j+m/k)+a]} \over {[b(j+m/k)+a]}}\right\}_{b=k^{-n}},  \eqno(1.16)$$
where $\gamma_1(a)$ is the first Stieltjes constant \cite{coffeyjmaa,coffey2009,stieltjes,wilton2}.
With (1.15) in hand, one may differentiate with respect to $s$ to find higher
order derivatives, and thus series representations for higher order Stieltjes
and other constants.  As a simple example, we have
\newline{\bf Corollary 7}.  For Re $s>0$, Re $a>0$, and $k\geq 2$ an integer, we
have             
$$\zeta''(s,a)-2{a^{1-s} \over {(s-1)^2}}\ln a-{{2a^{1-s}} \over {(s-1)^3}}-{{\ln^2 a} \over {2a^s}} - {a^{1-s} \over {s-1}}\ln^2 a$$
$$=-\sum_{n=0}^\infty {1 \over k^n}\sum_{j=0}^\infty \left\{{1 \over 2}\left({1 \over k}-1\right)\left[{{\ln^2(bj+a)} \over {(bj+a)^s}} + {{\ln^2(b(j+1)+a)} \over {(b(j+1)+a)^s}}\right]+{1 \over k}\sum_{m=1}^{k-1} {{\ln^2[b(j+m/k)+a]} \over {[b(j+m/k)+a]^s}}\right\}_{b=k^{-n}}.  \eqno(1.17)$$

The following cases for $\zeta'(s)$ could be presented as additional Corollaries.  
However, due to the celebrated case at $s=2$, for instance, for the 
Glaisher-Kinkelin constant $A$ \cite{glaisher,kinkelin} we state the following.
{\newline \bf Proposition 3}.  For Re $s>0$ we have (a) 
$$\zeta'(s)+{1 \over {(s-1)^2}}=-{1 \over 4}\sum_{n=0}^\infty {1 \over 2^n} 
\sum_{j=0}^\infty \left[{{2\ln[b(j+1/2)+1]} \over {[b(j+1/2)+1]^s}} -{{\ln(bj+1)}
\over {(bj+1)^s}}-{{\ln[b(j+1)+1}\over {[b(j+1)+1]^s}}\right]_{b=2^{-n}},  
\eqno(1.18a)$$
(b)
$$\zeta'(s)+{1 \over {(s-1)^2}}={1 \over 3}\sum_{n=0}^\infty {1 \over 3^n} 
\sum_{j=0}^\infty \left[{{\ln[b(j+1/3)+1]} \over {[b(j+1/3)+1]^s}} -{{\ln(bj+1)}
\over {(bj+1)^s}}-{{\ln[b(j+1)+1}\over {[b(j+1)+1]^s}}\right.$$
$$\left.+{{\ln[b(j+2/3)+1]} \over {[b(j+2/3)+1]^s}}\right]_{b=3^{-n}},  \eqno(1.18b)$$
and (c)
$$\zeta'(s)+{1 \over {(s-1)^2}}={1 \over 8}\sum_{n=0}^\infty {1 \over 4^n} 
\sum_{j=0}^\infty \left[{{2\ln[b(j+1/4)+1]} \over {[b(j+1/4)+1]^s}} -{{3\ln(bj+1)}
\over {(bj+1)^s}}-{{3\ln[b(j+1)+1}\over {[b(j+1)+1]^s}}\right.$$
$$\left.+{{2\ln[b(j+1/2)+1]} \over {[b(j+1/2)+1]^s}}+{{2\ln[b(j+3/4)+1]} \over {[b(j+3/4)+1]^s}} \right]_{b=3^{-n}}.  \eqno(1.18c)$$
We recall that $\ln A=-\zeta'(2)/\pi^2+[\ln(2\pi)+\gamma]/12$, or equivalently,
due to the functional equation of the Riemann zeta function, $\zeta'(-1)=1/12-
\ln A$.  Therefore, we have found a family of Addison-type series 
representations for $\ln A$.  Similarly from our results follow other 
Corollaries for constants such as $\zeta'(-2)=-\zeta(3)/4\pi^2$.  Other 
results for the Kinkelin constant $k=\zeta'(-1)$ are relegated to Appendix A.

The next Proposition describes that a vast array of constants may be written as
Addison-type series.  We have
{\newline \bf Proposition 4}.  Addison-type series may be developed for all of
the following quantities:
$$\int_0^x z^n \psi^{(m)}(a+bz)dz, ~~~~~~\int_0^x z^n \psi(a+bz)dz,~~~~~~
\int_0^x z^n \ln\Gamma(a+bz)dz.  \eqno(1.19)$$
Here, $n \geq 0$ and $m \geq 1$ are integers, and $a,b \in R$.
{\newline \bf Examples} of this Proposition include the constants
$$\int_0^{1/2} \ln \Gamma(z)dz={5 \over {24}}\ln 2+\ln(A^{3/2}\pi^{1/4}), \eqno(1.20a)$$
$$\int_0^1 z \ln \Gamma(z)dz=\ln\left({{(2\pi)^{1/4}} \over A}\right), \eqno(1.20b)$$
$$\int_0^1 z^2 \ln \Gamma(z)dz={1 \over 6}\ln(2\pi)-\ln A+{{\zeta(3)} \over {4\pi^2}}, \eqno(1.20c)$$
$$\int_0^{1/2} z \ln \Gamma(z)dz={1 \over {96}}\ln(16A^{24}\pi^6)-{7 \over {32}}
{{\zeta(3)} \over \pi^2},  \eqno(1.20d)$$
and
$$\int_0^{1/2} z^2 \ln \Gamma(z)dz={1 \over {5760}}[-55+62\ln 2+720\ln A+120\ln \pi
-(540/\pi^2)\zeta(3)-3600 \zeta'(-3)].  \eqno(1.20e)$$
An explicit example coming from Corollary 5 is:
{\newline \bf Corollary 6}.  We have
$$\int_0^1 \ln \Gamma(z)dz=\ln \sqrt{2\pi}={3 \over 4}-{1 \over 4}\sum_{n=0}^\infty
{1 \over 4^n}\sum_{j=0}^\infty \{j[\ln(bj+1)-\ln (bj)]+(j+1)[\ln(bj+b+1)-\ln (bj+b)]$$
$$-(2j+1)[\ln(2bj+b+2)-\ln (2bj+b)]\}_{b=2^{-n}}.  \eqno(1.21)$$

We let the hyperfactorial function \cite{bendersky} for $x \geq 0$
$$\ln K(x)={1 \over 2}(x^2-x)-{x \over 2}\ln 2\pi +\int_0^x \ln \Gamma(y)dy,
\eqno(1.22)$$
satisfying $K(0)=K(1)=K(2)=1$ and for $x>0$, $K(x+1)=x^x K(x)$.  Thus, for
nonnegative integers $n$, $K(n+1)=1^12^2\cdots n^n$.  In addition, we have the
value $K(1/2)=A^{3/2}/(2^{1/24}e^{1/8})$.  As follows from Proposition 4 we have
{\newline \bf Corollary 7}.  The hyperfactorial function $K$ has an integral
representation with $P_1$, and its values may be written as Addison-type series.

We introduce the generalized Somos constants \cite{somos} for $t>1$
$$\sigma_t = \prod_{n=1}^\infty n^{1/t^n}.  \eqno(1.23)$$
The Somos constant $\sigma_2 \simeq 1.66169$ has several representations,
including Ramanujan's nested radical expression.  We have
{\newline \bf Proposition 5}.  We have (a) the integral representations
$$\ln \sigma_t=\int_1^\infty {{\ln x} \over t^x}dx+\int_1^\infty\left({1 \over 
{x t^x}}-{{(\ln t) \ln x} \over t^x}\right)P_1(x)dx, \eqno(1.24a)$$
$$=\int_0^\infty\left[{e^{-x} \over {t-1}}+{1 \over {1-te^x}}\right]{{dx} \over x},
\eqno(1.24b)$$
and (b) the series formulas
$$\ln \sigma_t={1 \over {t-1}}\sum_{k=1}^\infty {{(-1)^{k-1}} \over k}
\mbox{Li}_k\left({1 \over t}\right)={1 \over {t-1}}\sum_{k=1}^\infty
{1 \over k}\left[t\mbox{Li}_k\left({1 \over t}\right)-1\right].  \eqno(1.25)$$
On the right side of (1.24a), we may identify
$\int_1^\infty t^{-x}\ln x ~dx=\Gamma(0,\ln t)/\ln t$, with $\Gamma(x,y)$ the
incomplete Gamma function and $\Gamma(0,x)=-\mbox{Ei}(-x)$, where Ei is the
exponential integral (e.g., \cite{nbs}, Ch. 5).  On the right side of (1.25), we have Li$_1(z)=-\ln(1-z)$.

{\bf Corollary 10}.  We have
$$\lim_{z\to 0^+}[\ln z+z\ln \sigma_{z+1}]=-\gamma.$$

Somos' quadratic recurrence is $g_n=ng_{n-1}^2$ with $g_0=1$.  We have
{\newline \bf Corollary 11}.  The solution $\ln g_n$ of Somos' quadratic
recurrence has an integral representation with $P_1$. 

Our methods directly extend to Dirichlet $L$ functions, as these may be written as a linear combination of Hurwitz zeta functions.  For instance, for $\chi$ a principal (nonprincipal) character modulo $m$ and Re $s \geq 1$ (Re $s \geq 0$) we have
$$L(s,\chi) = \sum_{k=1}^\infty {{\chi(k)} \over k^s} ={1 \over m^s}\sum_{k=1}^m 
\chi(k) \zeta\left(s,{k \over m}\right).  \eqno(1.26)$$
We detail the case of the Dirichlet $L$-function with a character modulo $4$ defined by
$$L(s) \equiv \sum_{m=0}^\infty {{(-1)^m} \over {(2m+1)^s}}, ~~~~~~~~~~~~
\mbox{Re} ~s>1, \eqno(1.27)$$
This function can be easily expressed as 
$$L(s)=4^{-s}[\zeta(s,1/4)-\zeta(s,3/4)]=1+4^{-s}[\zeta(s,5/4)-\zeta(s,3/4)], ~~~~
\mbox{Re} ~s \geq 0, \eqno(1.28)$$
and we have the particular values for nonnegative integers $m$ 
$$L(2m+1)=-{{(2\pi)^{2m+1}} \over {2(2m+1)!}}B_{2m+1}(1/4), \eqno(1.29)$$
where $B_m$ is the $m$th Bernoulli polynomial.  Generally the values of $L$
at odd or even integer argument may be expressed in terms of Euler or 
Bernoulli polynomials at rational argument and these in turn expressed in
terms of the Hurwitz zeta function.  Therefore we may in this way obtain many
Addison-type series representations for $L(2m)$ and $L(2m+1)$.  These include the
special cases of $L(1)=\pi/4$, $L(2)=G \simeq 0.91596559$, Catalan's 
constant, and $L(3)=\pi^3/32$.  

We have
{\newline \bf Proposition 6}.  For Re $s \geq 0$ we have
$$L(s)={1 \over 2}(1-3^{-s})+{1 \over 4}{1 \over {(s-1)}}(1-3^{1-s})$$
$$+4^{-s-1}\sum_{n=0}^\infty {1 \over 2^n} \sum_{j=0}^\infty \left[{1 \over {(bj+1/4)^s}}
-{2 \over {[b(j+1/2)+1/4]^s}}+{1 \over {[b(j+1)+1/4]^s}} \right.$$
$$\left.-{1 \over {(bj+3/4)^s}}+{2 \over {[b(j+1/2)+3/4]^s}}-{1 \over {[b(j+1)+3/4]^s}}\right]_{b=2^{-n}}.  \eqno(1.30)$$
This result itself gives many Corollaries.  Besides the above-mentioned examples of
$G=L(2)$ and $\pi/4=L(1)$, by differentiation we may obtain an Addison-type series
for $L'(s)=-4^{-s}(\ln 4)[\zeta(s,1/4)-\zeta(s,3/4)] +4^{-s}[\zeta'(s,1/4)-\zeta'(s,3/4)]$,
giving a series representation for the value
$$L'(1)={1 \over 4}\left\{(\ln 4)\left[\psi\left({3 \over 4}\right)-\psi\left({1 \over 4}\right)\right]+\gamma_1\left({3\over 4}\right)-\gamma_1\left({1\over 4}\right)\right\}$$
$$={1 \over 4}\left[-2\pi \ln 2+\gamma_1\left({3\over 4}\right)-\gamma_1\left({1\over 4}\right)\right],$$  
where we used the reflection formula of the digamma function.  In turn, we have
the value \cite{coffeystdiffs}
$$\gamma_1\left({3 \over 4}\right)-\gamma_1\left({1 \over 4}\right)=\pi\left\{\ln 8\pi+\gamma-2\ln\left[{{\Gamma\left({1 \over 4}\right)} \over {\Gamma\left({3 \over 4}\right)}}\right]\right\}. $$ 


We let for Re $s>1$ and $a \notin \{-1,-2,\ldots\}$
$$H(s,a) \equiv \sum_{n=1}^\infty {H_n \over {(n+a)^s}}.  \eqno(1.31)$$
There are at least two significant motivations for this function.  (i) it can
be analytically continued, and for $a=0$ can be expected to share some of the
properties of the derivative of the Riemann zeta function.  (ii)  At positive
integer values of $a$ it yields Euler sums (e.g., \cite{cofjcam1,flajolet}), and these have been of interest for some time.   
Simple examples include
$$\sum_{n=1}^\infty {H_n \over {(n+1)^2}}=\zeta(3), ~~~~~~~~\sum_{n=1}^\infty {H_n \over {(n+1)^3}}={1 \over 4}\zeta(4).  \eqno(1.32)$$
We have
\newline{\bf Proposition 7}.  We have
$$H(s,a)={1 \over {2(a+1)^s}}+\int_1^\infty {{[\psi(x+1)+\gamma]} \over {(x+a)^s}}
dx+\int_1^\infty \left[{{\psi'(x+1)} \over {(x+a)^s}}-s{{\psi(x+1)+\gamma]} \over
{(x+a)^{s+1}}}\right]P_1(x)dx.  \eqno(1.33)$$

We recall some relations useful in the following.

From the representation (cf. e.g., \cite{titch}, p. 14),
$$\zeta(s)={1 \over {s-1}}+{1 \over 2}-s\int_1^\infty x^{-(s+1)}P_1(x)dx, 
~~~~\mbox{Re} ~s >1, \eqno(1.34)$$
we obtain
$$\zeta^{(n)}(s)={{(-1)^n n!} \over {(s-1)^{n+1}}}+(-1)^n n\int_1^\infty {{P_1(x)}
\over x^{s+1}}\ln^{n-1} x ~dx-(-1)^n s \int_1^\infty {{P_1(x)}\over x^{s+1}}\ln^n x 
~dx.  \eqno(1.35)$$
Equation (1.34) extends to 
$$\zeta(s,a)={a^{-s} \over 2}+{a^{1-s} \over {s-1}}-s\int_0^\infty {{P_1(x)} \over
{(x+a)^{s+1}}} dx, ~~~~\mbox{Re} ~s >-1. \eqno(1.36)$$  

The defining Laurent expansion for the Stieltjes constants 
\cite{coffeyjmaa,coffey2009,stieltjes,wilton2} is 
$$\zeta(s,a)={1 \over {s-1}}+\sum_{k=0}^\infty {{(-1)^k \gamma_k(a)} \over
k!} (s-1)^k, ~~~~~~ s \neq 1, \eqno(1.37)$$
wherein $\gamma_0(a)=-\psi(a)$, and we have the connection to differences of logarithmic sums
$$\gamma_j(a)-\gamma_j(b)=\sum_{n=0}^\infty \left[{{\ln^j (n+a)} \over {n+a}}-
{{\ln^j (n+b)} \over {n+b}}\right], ~~~~~~j \geq 1,  \eqno(1.38)$$
where $a, b \in C/\{-1,-2,\ldots\}$.

\medskip
\centerline{\bf Proof of Propositions}

We include here proofs for the Propositions and Corollaries that are not
already covered in the previous section.

{\it Proposition 1}.  We apply a summation formula of \cite{titch} (p. 14) for
functions $\phi(x) \in C^1[\alpha,\beta]$:
$$\sum_{\alpha<n\leq \beta}\phi(n)=\int_\alpha^\beta \phi(x)dx+\int_\alpha^\beta \phi'(x)P_1(x)dx +P_1(\alpha)\phi(\alpha)-P_1(\beta)\phi(\beta).  \eqno(2.1)$$
We take $\phi(x)=z^x/(x+a)^s$ and $\alpha$ and $\beta$ to be positive integers,
so that $P_1(\alpha)=P_1(\beta)=-1/2$.  Then we have
$$\sum_{n=\alpha+1}^\beta {z^n \over {(n+a)^s}}=\int_\alpha^\beta {z^x \over
{(x+a)^s}} dx+\int_\alpha^\beta \left[{{z^x \ln z} \over {(x+a)^s}}-{{sz^x} \over
{(x+a)^{s+1}}}\right]P_1(x)dx$$
$$+{1 \over 2}\left[{z^\beta \over {(\beta+a)^s}}-{z^\alpha \over {(\alpha+a)^s}}\right].  \eqno(2.2)$$
In order to obtain the Proposition we take $\alpha=1$, $\beta\to \infty$,
and add to both sides $1/a^s+z/(a+1)^s$, using
$$\sum_{n=2}^\infty {z^n \over {(n+a)^s}} + {1 \over a^s}+{z \over {(a+1)^s}}
=\Phi(z,s,a).  \eqno(2.3)$$

{\it Corollary 1}.  For the Corollary we put $a=1$ in (1.4), make a simple
change of variable, and use the $1$-periodicity of $P_1$, finding
$$\Phi(z,s,1)=1+ {z \over 2^{s+1}}+\int_2^\infty {z^{y-1} \over y^s}dy +\int_2^\infty\left[{{z^{y-1} \ln z} \over y^s}-{{sz^{y-1}} \over y^{s+1}}\right]
P_1(y)dy$$
$$=1+ {z \over 2^{s+1}}+\int_1^\infty {z^{y-1} \over y^s}dy +\int_1^\infty\left[{{z^{y-1} \ln z} \over y^s}-{{sz^{y-1}} \over y^{s+1}}\right]
P_1(y)dy$$
$$-\int_1^2 {z^{y-1} \over y^s}dy -\int_1^2\left[{{z^{y-1} \ln z} \over y^s}-{{sz^{y-1}} \over y^{s+1}}\right]\left(y-{3 \over 2}\right)dy.  \eqno(2.4)$$
The second line of this equation, using integration by parts, evaluates to $-z^{y-1}y^{-s}(y-3/2)|_{y=1}^{y=2}=-1/2-z/2^{s+1}$, giving the Corollary.

{\it Corollary 3}.  By using Corollary 1 and interchanging integrations we have
$$\int_0^1 t^{\alpha-1} \mbox{Li}_s(t)dt={1 \over 2}{1 \over {(\alpha+1)}}
+\int_1^\infty {1 \over x^s}{{dx} \over {(x+\alpha)}}-\int_1^\infty \left[
{1 \over x^s}{1 \over {(x+\alpha)^2}}-{s \over x^{s+1}}{1 \over {(x+\alpha)}} \right]P_1(x)dx$$
$$={1 \over 2}{1 \over {(\alpha+1)}}+{1 \over s} ~_2F_1(1,s;s+1;-\alpha)
-\int_1^\infty \left[{1 \over x^s}{1 \over {(x+\alpha)^2}}-{s \over x^{s+1}}{1 \over {(x+\alpha)}} \right]P_1(x)dx. \eqno(2.5)$$
Therefore, recalling that $_2F_1(1,1;2;-\alpha)=\ln(1+\alpha)/\alpha$ and using its
derivative with respect to $\alpha$, at $s=2$ we obtain
$$\int_0^1 t^{\alpha-1} \mbox{Li}_2(t)dt
={1 \over 2}{1 \over {(\alpha+1)}}+{1 \over \alpha}-{1 \over \alpha^2} \ln(\alpha+1) -\int_1^\infty {{P_1(x)dx} \over {x^2(x+\alpha)^2}}-2\int_1^\infty {{P_1(x)dx} \over {x^3(x+\alpha)}}.  \eqno(2.6)$$
We then use partial fractions for the integrals on the right side of this
equation, together with relation (1.34):
$$\int_1^\infty {{P_1(x)dx} \over {x^2(x+\alpha)^2}}+2\int_1^\infty {{P_1(x)dx} \over {x^3(x+\alpha)}}=\int_1^\infty \left[{2 \over {\alpha x^3}}
-{1 \over {\alpha^2 x^2}}+{1 \over \alpha^2}{1 \over {(x+\alpha)^2}}\right]P_1(x)
dx$$
$$={2 \over \alpha}\left[{3 \over 4}-{{\zeta(2)} \over 2}\right]+{1 \over \alpha^2}
\left(\gamma-{1 \over 2}\right)+{1 \over \alpha^2}\int_1^\infty {{P_1(x)dx} \over
{(x+\alpha)^2}}.  \eqno(2.7)$$
The proof of (a) is completed by using the representation (e.g., \cite{coffeyjnt} (2.20)) 
$$\psi(a+1)=\ln(a+1)-{1 \over {2(a+1)}}+\int_1^\infty {{P_1(y)dy} \over {(y+a)^2}}
.  \eqno(2.8)$$

For part (b), we note that the function $h(s)=\int_1^\infty x^{-(s+1)}P_1(x)dx$
comes from (1.34).  We make use of (2.5) together with the partial fractions
decomposition
$${1 \over {x^n(x+\alpha)^2}}+{n \over {x^{n+1}(x+\alpha)}}=(-1)^n\left[{1 \over {\alpha^n (x+\alpha)^2}}+\sum_{j=1}^n {{(-1)^j j} \over {\alpha^{n-j+1}x^{j+1}}}\right].  
\eqno(2.9)$$
Using again the integral representation (2.8), the rest of the Corollary follows.

Remarks.  Integrations or differentiations of (2.9) yield families of other
relations.  The function of (1.7b) may be written as 
$${1 \over n} ~_2F_1(1,n;n+1;-\alpha)={{(-1)^{n+1}} \over \alpha^n}\ln(1+\alpha)
-\sum_{j=1}^{n-1} {{(-1)^j} \over {(n-j)\alpha^j}}.  \eqno(2.10)$$
The $_2F_1$ function of (2.5) is transformable as
$${1 \over s} ~_2F_1(1,s;s+1;-\alpha)={1 \over s}{1 \over {(1+\alpha)}} ~_2F_1\left
( 1,1;s+1;{\alpha \over {1+\alpha}}\right).  \eqno(2.11)$$
We have the special cases
$$\int_0^1 t^{\alpha-1}\mbox{Li}_1(t)dt=-\int_0^1 t^{\alpha-1}\ln(1-t) ~dt
={1 \over \alpha}[\psi(\alpha+1)+\gamma], ~~~~~\mbox{Re} ~\alpha>0, \eqno(2.12)$$
and
$$\int_0^1 {1 \over t}\mbox{Li}_s(t)dt=\zeta(s+1), ~~~~~\mbox{Re} ~s>0.  \eqno(2.13)$$
The latter may be easily verified by term-by-term integration of the series
form of Li$_s$.

{\it Proposition 2}.  We have from (1.36) that
$$\zeta'(s,a)=-{{\ln a} \over {2a^s}}-{a^{1-s} \over {(s-1)^2}}-{{a^{1-s}\ln a}
\over {(s-1)}}-\int_0^\infty {{P_1(x)dx} \over {(x+a)^{s+1}}}
+s\int_0^\infty {{P_1(x)\ln(x+a)} \over {(x+a)^{s+1}}}dx.  \eqno(2.14)$$
We make use of the functions for $k \geq 2$ and $f(x)=-P_1(x)$ \cite{coffeyjnt},
$$g_k(x)=f(x)-{1 \over k}f(kx),  \eqno(2.15)$$
with $\sum_{n=0}^\infty {{g_k(k^n x)} \over k^n}=f(x)$.  
Then
$$\zeta'(s,a)+{{\ln a} \over {2a^s}}+{a^{1-s} \over {(s-1)^2}}+{{a^{1-s}\ln a}
\over {(s-1)}}$$
$$=\sum_{n=0}^\infty {1 \over k^n}\int_0^\infty {{[1-s\ln(x+a)]} \over {(x+a)^{s+1}
}}g_k(k^nx)dx$$
$$=\sum_{n=0}^\infty {1 \over k^{2n}}\int_0^\infty {{[1-s\ln(k^{-n}y+a)]} \over {(k^{-n}y+a)^{s+1} }}g_k(y)dy$$
$$=\sum_{n=0}^\infty {1 \over k^{2n}}\sum_{j=0}^\infty \int_j^{j+1} {{[1-s\ln(k^{-n}y+a)]} \over {(k^{-n}y+a)^{s+1} }}g_k(y)dy$$
$$=\sum_{n=0}^\infty {1 \over k^{2n}}\sum_{j=0}^\infty \left[{1 \over 2}\left(1-
{1 \over k}\right)\int_j^{j+1/k} + {1 \over 2}\left(1-{3 \over k}\right) \int_{j+1/k}^{j+2/k} + {1 \over 2}\left(1- {5 \over k}\right)\int_{j+2/k}^{j+3/k} 
\right.$$
$$\left. + \ldots + {1 \over 2}\left({1 \over k}-1\right)\int_{j+(k-1)/k}^{j+1} 
\right]{{[1-s\ln(k^{-n}y+a)]} \over {(k^{-n}y+a)^{s+1} }}g_k(y)dy.  \eqno(2.16)$$
In the last step we have used the values of $g_k(x)$ on subintervals $\left[{{j-1} \over k},{j \over k} \right)$ for $j=1,2,\ldots,k$, 
$$g_k(x)={1 \over 2}\left(1-{1 \over k}\right)-{{(j-1)} \over k}, ~~~~~~x \in \left[{{j-1} \over k},{j \over k}\right).  \eqno(2.17)$$
In particular, the difference of these values on consecutive subintervals is
simply $1/k$.  Performing the integrations and collecting the terms gives the Proposition.

Remark.  Let us consider a general relation coming from (2.1).  Suppose that
$\alpha=0$ there, that $\phi(\beta)P_1(\beta) \to 0$ as $\beta \to \infty$, and
that $\int_0^\infty \phi(x)dx$ converges.
Then employing the $g_k$ functions of (2.15) we have
$$\sum_{n=1}^\infty \phi(n)=\int_0^\infty \phi(x)dx+\int_0^\infty \phi'(x)P_1(x)dx
-{1 \over 2}\phi(0)$$
$$=\int_0^\infty \phi(x)dx-\sum_{n=0}^\infty {1 \over k^n}\int_0^\infty \phi'(x)
g_k(k^nx)dx-{1 \over 2}\phi(0)$$
$$=\int_0^\infty \phi(x)dx-\sum_{n=0}^\infty {1 \over k^{2n}}\int_0^\infty \phi'(
k^{-n}y)g_k(y)dy-{1 \over 2}\phi(0). \eqno(2.18)$$
Then given the piecewise values (2.17) of $g_k$, the $g_k$ term on the right
side can always be written as an explicit summation.

{\it Proposition 4}.  That the stated integrals may be explicitly evaluated
follows from Theorems 4.2, 4.3, and 4.5 of \cite{moll}. 
Then from Corollary 5, the relation below (1.13) for $\psi^{(n)}(x)$, and
Proposition 2, the Proposition follows by a corresponding integration.

Remark.  In the Appendix, we discuss the prevalent functions $A_k(q)$ of \cite{moll}, writing them in terms of sums of the Stieltjes constants, and delineating some of their properties in that manner.

{\it Proposition 5}.  We have
$$\ln \sigma_t=\sum_{n=2}^\infty {{\ln n} \over t^n}=-\left. {{\partial \mbox{Li}_s(1/t)} \over {\partial s}}\right|_{s=0}.  \eqno(2.19)$$
As from Corollary 1,
$$-{{\partial \mbox{Li}_s(z)} \over {\partial s}}=\int_1^\infty {z^x \over x^s}\ln x ~dx +\int_1^\infty \left({z^x \over x^s} (\ln z)\ln x+{z^x \over x^{s+1}}-s {z^x \over x^{s+1}} \ln x\right)P_1(x)dx, \eqno(2.20)$$
we obtain part (1.24a).  For (1.24b), we use for Re $z>0$
$$\ln z=\int_0^\infty (e^{-t}-e^{-tz}){{dt} \over t}, \eqno(2.21)$$
and then interchange summation and integration to write
$$\ln \sigma_t=\sum_{n=2}^\infty {{\ln n} \over t^n}=\sum_{n=2}^\infty {1 \over t^n}
\int_0^\infty (e^{-x}-e^{-nx}){{dx} \over x}.  \eqno(2.22)$$

For part (b) we reorder a double sum as a first demonstration,
$$\sum_{k=1}^\infty {{(-1)^{k-1}} \over k}\mbox{Li}_k\left({1 \over t}\right)=
\sum_{n=1}^\infty {1 \over t^n} \sum_{k=1}^\infty {{(-1)^{k-1}} \over k} {1 \over
n^k}=\sum_{n=1}^\infty {1 \over t^n}\ln \left(1+{1 \over n}\right)$$
$$=\sum_{n=1}^\infty {1 \over t^n}[\ln (n+1)-\ln n]=(t-1)\ln \sigma_t.  \eqno(2.23)$$
The other summation in (1.25) is obtained similarly.

For a second proof we appeal to part (a).  From Corollary 1 we obtain
$$\sum_{k=1}^\infty {{(-1)^{k-1}} \over k}\mbox{Li}_k\left({1 \over t}\right)
={{\ln 2} \over {2t}}+\int_1^\infty {1 \over t^x}\ln\left({{x+1} \over x}\right)dx$$
$$+\int_1^\infty \left[{{\ln(1/t)} \over t^x}\ln\left({{x+1} \over x}\right)-{1 \over t^x}{1 \over {x(x+1)}}\right]P_1(x)dx.  \eqno(2.24)$$
We use the property
$$\int_1^\infty {{f(x+1)} \over t^x}dx=t\int_2^\infty {{f(y)} \over t^y}dy=
t\int_1^\infty {{f(y)} \over t^y}dy -t\int_1^2 {{f(y)} \over t^y}dy,  \eqno(2.25)$$
giving
$$\sum_{k=1}^\infty {{(-1)^{k-1}} \over k}\mbox{Li}_k\left({1 \over t}\right)
={{\ln 2} \over {2t}}+(t-1)\int_1^\infty {{\ln x} \over t^x}dx-t\int_1^2 {{\ln x}
\over t^x}dx$$
$$+(t-1)\int_1^\infty\left[{{\ln(1/t)} \over t^x} \ln x-{1 \over {xt^x}}\right]
P_1(x)dx-t\int_1^2{{\ln(1/t)} \over t^x} \ln x P_1(x)dx-t\int_1^2 {{P_1(x)} \over
{xt^x}}dx.  \eqno(2.26)$$
By using integration by parts, it is seen that the three integrals on $[1,2)$,
wherein $P_1(x)=x-3/2$, cancel the $(\ln 2)/(2t)$ term, and the summation
formula follows.

{\it Corollary 10}.  We may use (1.24a), (1.24b), or (1.25).  Using (1.25) we have
$$\ln z+z\ln \sigma_{z+1}=\ln z+\sum_{k=1}^\infty {{(-1)^{k-1}} \over k}\mbox{Li}_k
\left({1 \over {z+1}}\right).  \eqno(2.27)$$
As $z \to 0^+$, the logarithmic singularity is cancelled since
Li$_1[1/(1+z)]=-\ln z+\ln (1+z)$.  Therefore, we obtain
$$\lim_{z\to 0^+}[\ln z+z\ln \sigma_{z+1}]=\sum_{k=2}^\infty {{(-1)^{k-1}} \over k}
\zeta(k)=\int_0^\infty \left({e^{-t} \over t}+{1 \over {1-e^{-t}}}\right)dt =-\gamma. \eqno(2.28)$$

Alternatively, from (1.24a), we have
$$\ln z+z \ln \sigma_{z+1}=\ln z + z\left\{-{{\mbox{Ei}[-\ln(1+z)]} \over {\ln(1+z)
}}+\int_1^\infty\left({1 \over {x t^x}}-{{(\ln t) \ln x} \over t^x}\right)P_1(x)dx
\right\}$$
$$=\ln z+z\left[-(\gamma+\ln z){1 \over z}+{1 \over 2}(3-\gamma-\ln z)+O(z)\right]$$
$$=-\gamma + O(z \ln z),  \eqno(2.29)$$
and the Corollary follows again.

Remarks.  The series of (2.28) is an example of the expansion \cite{grad} (p. 939)
$$\ln \Gamma(x)=-\ln x-\gamma x+\sum_{k=2}^\infty (-1)^k {{\zeta(k)} \over k}x^k,
~~~~~~|x|<1. \eqno(2.30)$$  
Similarly using the second equality of (1.25) gives the sum
$$-1+\sum_{k=2}^\infty {{[\zeta(k)-1]} \over k}=-\gamma.$$

{\it Corollary 11}.  The solution of Somos's quadratic recurrence is given by
\cite{somos}
$$\ln g_n=2^n \ln \sigma_2+{1 \over 2}{{\partial \Phi} \over {\partial s}}\left(
{1 \over 2},0,n+1\right).  \eqno(2.31)$$
From Proposition 1 we have
$${{\partial \Phi} \over {\partial s}}\left(z,s,a\right)=-{{\ln a} \over a^s}-{z
\over 2}{{\ln(a+1)} \over {(a+1)^s}}-\int_1^\infty {{z^x \ln(x+a)} \over {(x+a)^s}}
dx$$
$$-\int_1^\infty \left[{{z^x(\ln z)} \over {(x+a)^s}} \ln(x+a)  +{z^x \over {(x+a)^{s+1}}} + s {{z^x \ln (x+a)} \over {(x+a)^{s+1}}}\right]P_1(x)dx,
\eqno(2.32)$$
giving
$${{\partial \Phi} \over {\partial s}}\left({1 \over 2},0,n+1\right)=-\ln(n+1)
-{1 \over 4}\ln(n+2)-\int_1^\infty {{\ln(x+n+1)} \over 2^x}dx$$
$$+\int_1^\infty \left[\ln 2 {{\ln(x+n+1)} \over 2^x}-{1 \over 2^x}{1 \over {(x+
n+1)}}\right]P_1(x)dx.  \eqno(2.33)$$
Together with the representation for $\ln \sigma_2$ given in Proposition 5,
this completes the Corollary.

{\it Remark}.  The result of this Corollary easily extends to the generalized
Somos recurrence $g_n=ng_{n-1}^t$ with $g_0=1$ and $n \geq 1$.  For the 
solution is given by
$$\ln g_n=t^n \ln \sigma_t+{1 \over t}{{\partial \Phi} \over {\partial s}}\left(
{1 \over t},0,n+1\right),  \eqno(2.34)$$
and Proposition 1 again applies.

{\it Proposition 6}.  Using (1.36) we have
$$L(s)={1 \over 4^s}\left\{{4^s \over 2}(1-3^{-s})+{4^{s-1} \over {(s-1)}}(1-3^{1-s})
\right.$$
$$\left.-s\int_0^\infty \left[{1 \over {(x+1/4)^{s+1}}}-{1 \over {(x+3/4)^{s+1}}}\right]
P_1(x)dx \right\}.  \eqno(2.35)$$
We then employ the case of the function $g_2(x)$ of (2.15) in order to evaluate
the integrals.  We omit further details.

Remark.  We may extend Proposition 6 with the use of any of the functions $g_k(x)$.

{\it Proposition 7}.  We apply (2.1) with $\phi(x)=[\psi(x+1)+\gamma]/(x+a)^s$
and $\phi(n)=H_n/(n+a)^s$.  

Remark.  This Proposition may be extended by using the generalized harmonic
numbers (1.14).

\bigskip
\centerline{\bf Summary}
\medskip

We have obtained integral representations for a wide variety of special
functions, including the Lerch zeta, the polylogarithm, Dirichlet $L$- and the Clausen functions.  Furthermore, we have shown that the method of Addison may be generalized in several directions to develop series representations for an
array of mathematical constants.  These constants include, but are not limited to,
those that arise from definite integrations of log Gamma, digamma, and polygamma
functions, with or without powers, and those that arise from certain 
derivatives of the Lerch zeta, polylogarithm, and Hurwitz zeta functions.
An instance of the latter category is that for the generalized Somos constants
$\sigma_t$.  By using the stepwise functions $g_k(x)$ in the Addison approach, general series representations with parameter $k$ are obtained.  Our methods have
broad applicability and permit both the recovery of known results and the
development of novel integral and series representations.



\bigskip
\centerline{\bf Appendix A: on Kinkelin's constant $k$ and $\Gamma$ function integrals}
\medskip

Here we present material on Kinkelin's constant $k=\zeta'(-1) \simeq -0.165421$
together with some particular integrals with the Gamma function.  Several expressions for the Kinkelin constant are known \cite{kinkelin,sri}.  For instance,
$$k=2\int_0^\infty {{x \ln x} \over {e^{2\pi x}-1}}dx, \eqno(A.1)$$
and expanding the denominator as a geometric series verifies the relation
$k=1/12-\ln A$.  Another relation is  
$$k={1 \over {12}}-{1 \over 4}\ln (2\pi)+\int_0^1 x\Gamma(x)dx.  \eqno(A.2)$$
Introducing the constants $c_k$ in $\Gamma(z+1)=\sum_{k=0}^\infty c_k z^k$
for $|z|<1$, with $c_0=1$ and $c_1=-\gamma$, there is the recurrence 
$c_{n+1}={1 \over {n+1}}\sum_{k=0}^n (-1)^{k+1} s_{k+1}c_{n-k}$, where
$s_1=\gamma$ and $s_n=\zeta(n)$ for $n \geq 2$ \cite{grad} (p. 935).  We 
then have 
{\newline \bf Proposition A1}.  We have
$$\int_0^1 x\Gamma(x)dx={1 \over 2}+{\gamma \over 6}+\sum_{k=2}^\infty {{(-1)^k
c_k} \over {(k+1)(k+2)}}=1-{\gamma \over 2}+\sum_{k=2}^\infty {c_k \over {k+1}}.
\eqno(A.3)$$

{\it Proof}.  For the first equality we employ the Taylor expansion of $\Gamma$
about $x=1$, and then evaluate a simple Beta function integral:
$$\int_0^1 x\Gamma(x)dx=\sum_{k=0}^1 (-1)^k c_k \int_0^\infty x(1-x)^k dx
=\sum_{k=0}^\infty {{(-1)^k c_k} \over {(k+1)(k+2)}}.  \eqno(A.4)$$
For the second equality we identify the integral with $\int_0^1 \Gamma(x+1)dx$
and proceed similarly.

We have
{\newline \bf Proposition A2}.  We have
$$k=-{\gamma \over {12}}+\ln 2-{5 \over 6}+{1 \over 2}\sum_{n=2}^\infty
{{(-1)^n[\zeta(n)-1]} \over {(n+1)(n+2)}}.  \eqno(A.5)$$

{\it Proof}.  We reorder the double series
$$\sum_{n=2}^\infty {{(-1)^n} \over {(n+1)(n+2)}} \sum_{j=2}^\infty {1 \over j^n}
=\sum_{j=2}^\infty \sum_{n=2}^\infty {{(-1)^n} \over {(n+1)(n+2)}}{1 \over j^n}
=\sum_{j=2}^\infty \sum_{n=2}^\infty (-1)^n\left({1 \over {n+1}}-{1 \over {n+2}}
\right){1 \over j^n}$$
$$=\sum_{j=2}^\infty \left[{{6j^2+3j-1} \over {6j}}+j(j+1)\ln\left({{j+1} \over j}
\right)\right]$$
$$={{11} \over 6}+{\gamma \over 6}-2\ln A-2\ln 2.  \eqno(A.6)$$
The latter expression may be found by using the limit relations $\gamma=\lim_{N\to \infty}\left(\sum_{k=1}^n {1 \over k}-\ln n\right)$ and for the Glaisher-Kinkelin constant
$$\ln A=\lim_{N\to \infty}\left[\sum_{k=1}^n k \ln k-\left({N^2 \over 2}+{N \over 2
}+{1 \over {12}}\right)\ln N+{N^2 \over 4}\right].  \eqno(A.7)$$
Using the relation $k=1/12-\ln A$ completes the Proposition.

The alternating sum in (A.5) also occurs in \cite{choisri} (p. 116), where it is calculated with another approach.  Corollaries of Proposition A2 follow by using various integral representations of $\zeta(n)$ in (A.5). For instance, by using
(1.26) we have
{\newline \bf Corollary A1}.  We have
$$k={1 \over 6}\ln 2-{5 \over {18}}-{1 \over 2}\int_1^\infty \left[2-(2x+1)\ln
\left({{x+1} \over x}\right)\right]P_1(x)dx.  \eqno(A.8)$$
In this expression, the integral term provides a small correction (approximately
2\%).  This Corollary enables other Addison-type series to be developed for
Kinkelin's constant.

Expecting that the values of $\int_0^1 x\Gamma(x)dx \approx 0.92746$ and
$\int_0^1 \sin x ~\Gamma(x)dx \approx 0.872427$ are not too different, we have
performed a brief investigation of the following integrals.  We have
{\newline \bf Proposition A3}.  For $\lambda>0$ and $|\alpha| \leq \pi/2$ we have the representation
$$\int_0^1 \sin (\alpha x)\Gamma(x)dx=\int_0^\infty e^{-\lambda t\cos \alpha}
\sin(\lambda t \sin \alpha){{(\lambda t-1)} \over {t(\ln t+\ln \lambda)}}dt.
\eqno(A.9)$$

{\it Proof}.  This follows by using an integral representation for 
$\Gamma(x)\sin \alpha x$ \cite{grad} (p. 935) and interchanging integrations.

From A3 we deduce
{\newline \bf Corollary A2}.  For $\lambda>0$ we have
$$\int_0^1 x\Gamma(x)dx=\lambda \int_0^\infty e^{-\lambda t}{{(\lambda t-1)} \over
{\ln(\lambda t)}}dt =\int_0^\infty e^{-u}{{(u-1)} \over {\ln u}}du.  \eqno(A.10)$$

{\it Proof}.  We let $\alpha \to 0$ in (A.9) and use
$e^{-\lambda t\cos \alpha} \sin(\lambda t \sin \alpha)=\alpha \lambda te^{-\lambda t}+O(\alpha^3)$.

We now give a direct proof of this Corollary.  We interchange integrations 
(justified by absolute convergence) and integrate by parts twice:
$$\int_0^1 x\Gamma(x)dx=\int_0^1\int_0^\infty e^{-t}x t^{x-1}dtdx$$
$$=\int_0^\infty e^{-t}\left({1 \over {t\ln^2 t}}-{1 \over {\ln^2 t}}+{1 \over 
{\ln t}}\right)dt$$
$$=-\int_0^\infty {e^{-t} \over {\ln^2 t}}dt=\int_0^\infty e^{-t}{{(t-1)} \over {\ln t}}dt.  \eqno(A.11)$$

The next result recovers A1 as a special case.  Again, $_pF_q$ is the generalized hypergeometric function.  We have
{\newline \bf Proposition A4}.  We have for $|\alpha| \leq \pi/2$
$$\int_0^1 \sin(\alpha x)\Gamma(x)dx=\alpha \sum_{k=0}^\infty {{(-1)^k c_k} \over
{(k+1)(k+2)}} ~_1F_2\left(1;{{k+3} \over 2}, 2+{k \over 2};-{\alpha^2 \over 4}
\right)$$  
$$={{1-\cos \alpha} \over \alpha}+{\gamma \over \alpha^2}(\alpha-\sin \alpha)+\alpha \sum_{k=2}^\infty {{(-1)^k c_k} \over
{(k+1)(k+2)}} ~_1F_2\left(1;{{k+3} \over 2}, 2+{k \over 2};-{\alpha^2 \over 4}
\right)\eqno(A.12)$$
$$=\sum_{k=0}^\infty (-1)^k c_k\left[\sum_{\ell=0}^k {{\ell!} \over \alpha^{\ell+1}}{k \choose \ell} \cos\left({{\ell
\pi} \over 2}\right)-{{k!} \over \alpha^{k+1}}\cos\left(\alpha-{{k \pi} \over 2}
\right)\right]$$
$$=\sum_{k=0}^\infty (-1)^k c_k\left[\sum_{\ell=0}^{k-1} {{\ell!} \over \alpha^{\ell+1}}{k \choose \ell} \cos\left({{\ell \pi} \over 2}\right)+2{{k!} \over \alpha^{k+1}}\sin\left({\alpha \over 2}\right)\sin\left({{\alpha-k \pi} \over 2}\right)\right].  \eqno(A.13)$$

{\it Proof}.  The Taylor expansion of $\Gamma$ about $x=1$ is first used to write
$$\int_0^1 \sin(\alpha x)\Gamma(x)dx=\sum_{k=0}^\infty c_k \int_0^1 \sin \alpha x
(x-1)^kdx$$
$$=\sum_{k=0}^\infty (-1)^k c_k\sum_{j=0}^\infty {{(-1)^j} \over {(2j+1)!}} 
\alpha^{2j+1} \int_0^1 x^{2j+1} (1-x)^k dx$$
$$=\sum_{k=0}^\infty (-1)^k c_k\sum_{j=0}^\infty (-1)^j \alpha^{2j+1}
{{\Gamma(k+1)} \over {\Gamma(2j+k+3)}} {{(1)_j} \over {j!}}, \eqno(A.14)$$
where the integral evaluated in terms of the Beta function $B$ as $B(2j+2,k+1)$,
and $(z)_j=\Gamma(z+j)/\Gamma(z)$ is the Pochhammer symbol.
We use Legendre's duplication formula for the Gamma function to write
$\Gamma(2j+k+3)={1 \over \sqrt{\pi}}2^{2j+k+2}\Gamma\left(j+{{k+3} \over 2}\right)
\Gamma\left(j+{k \over 2}+2\right)$ and $\Gamma(k+3)={1 \over \sqrt{\pi}} 2^{k+2}\Gamma\left({{k+3} \over 2}\right)\Gamma\left({k \over 2}+2\right)$.
These relations yield
$$\int_0^1 \sin(\alpha x)\Gamma(x)dx=\sum_{k=0}^\infty c_k {{\Gamma(k+1)} \over
{\Gamma(k+3)}}\sum_{j=0}^\infty {{(-1)^j \alpha^{2j+1}} \over 2^{2j}} {1 \over
{\left({{k+3} \over 2}\right)_j\left({k \over 2}+2\right)_j}} {{(1)_j} \over {j!}}$$
$$=\alpha \sum_{k=0}^\infty {{(-1)^k c_k} \over {(k+1)(k+2)}} ~_1F_2\left(1;{{k+3} \over 2}, 2+{k \over 2};-{\alpha^2 \over 4}\right).  \eqno(A.15)$$  

For the alternative form (A.13) we write the first integral on the right side of (A.14) as 
$$\int_0^1 \sin \alpha x (1-x)^kdx=\int_0^1 \sin[\alpha(1-v)] v^k dv$$
$$=\int_0^1 [\sin \alpha \cos \alpha v-\sin \alpha v\cos \alpha]v^kdv$$
$$=\sin \alpha\sum_{\ell=0}^k \ell!{k \choose \ell} {v^{k-\ell} \over \alpha^{\ell
+1}} \left.\sin \left(\alpha v +{{\ell \pi} \over 2}\right)\right|_{v=0}^{v=1}
+\cos \alpha\sum_{\ell=0}^k \ell!{k \choose \ell} {v^{k-\ell} \over \alpha^{\ell
+1}} \left.\cos \left(\alpha v +{{\ell \pi} \over 2}\right)\right|_{v=0}^{v=1},
\eqno(A.16)$$
where \cite{grad} (pp. 183-184) were used.  We then find
$$\int_0^1 \sin \alpha x (1-x)^kdx=\sum_{\ell=0}^k \ell!{k \choose \ell}\left[\sin \alpha
\sin\left(\alpha+{{\ell \pi} \over 2}\right)+\cos \alpha \cos\left(\alpha+{{\ell \pi} \over 2}\right)\right]{1 \over \alpha^{\ell+1}}$$
$$-{{k!} \over \alpha^{k+1}}\left[\sin \alpha \sin\left({{k \pi} \over 2}\right)
+\cos \alpha \cos\left({{k \pi} \over 2}\right)\right], \eqno(A.17)$$
leading to (A.13).

Remark.  When $\alpha \to 0$, $_1F_2 \to 1$ in (A.12) and (A.3) results.


\pagebreak
\centerline{\bf Appendix B: The functions $A_k(q)$ of \cite{moll}}
\medskip

The functions for integers $k$ 
$$A_k(q) \equiv k \left.{\partial \over {\partial z}}\zeta(z,q)\right|_{z=1-k},
\eqno(B.1)$$
are very useful in evaluating integrals over the Hurwitz zeta function
\cite{moll}.  We present a study of these functions, first writing a sum
representation for them in terms of the Stieltjes constants.  Since much other
information is known about these functions, including their connection with
negapolygamma functions, we effectively unify many descriptions of these sets
of functions.  We show that it is possible to recover known properties \cite{moll} of the $A_k(q)$ functions via their representation in terms of the Stieltjes constants.  Then we write an integral representation for these functions when $k<2$, and mention
two summation relations.  Here we take $0 \leq q \leq 1$.

{\bf Lemma B1}.  We have 
$$A_k(q)=-{1 \over k}-k\sum_{n=0}^\infty {{\gamma_{n+1}(q)} \over {n!}}k^n, 
\eqno(B.2)$$
and 
$$A_1(q)=\ln \Gamma(q)-{1 \over 2}\ln 2\pi=-1-\sum_{n=0}^\infty {{\gamma_{n+1}(q)} \over {n!}}.  \eqno(B.3)$$

{\bf Lemma B2}.  The relations $\int_0^1 A_k(q)dq=0$ and
$$\sum_{n=0}^\infty {k^n \over {n!}}\int_0^1 \gamma_{n+1}(q)dq=-{1 \over k^2}
\eqno(B.4)$$
are equivalent.

{\bf Lemma B3}.  We have $A_k(0)=A_k(1)=k\zeta'(1-k)$ for $k \geq 2$.

Let $B_k(q)=-k\zeta(1-k,q)$ be the Bernoulli polynomials.  Then we have
\newline{\bf Lemma B4}.  We have
$$B_k(q)=1-k\sum_{n=0}^\infty {{\gamma_n(q)} \over {n!}} k^n.  \eqno(B.5)$$
{\newline \bf Corollary B1}.  We have for the Bernoulli numbers $B_k=B_k(0)$
$$(-1)^kB_k=B_k(1)=1-k\gamma-k\sum_{n=1}^\infty {\gamma_n \over {n!}}k^n, 
\eqno(B.6)$$
and in particular at $k=1$ we recover the special case
$$\sum_{n=1}^\infty {\gamma_n \over {n!}}={1 \over 2}-\gamma.  \eqno(B.7)$$

{\bf Lemma B5}.  We have $A_{k+1}(q)=A_k(q)+kq^{k-1} \ln q$ and thus for
integers $n \geq 0$
$$A_{k+n}(q)=A_k(q)+\ln q \sum_{j=0}^{n-1}(k+j)q^{k+j-1}.  \eqno(B.8)$$

{\bf Lemma B6}.  We have
$$A'_{k+1}(q)=(k+1)\left[A_k(q)+{1 \over k}B_k(q)\right].  \eqno(B.9)$$

Proof of Lemmas.  {\it Lemma B1}.  We use the expansion (1.37) and the definition
(B.1).  The relation (B.3) corresponds to Corollary 5 of \cite{coffey2009}.

{\it Lemma B2}.  We use Proposition 4 of \cite{coffeyjmaa}, differentiating 
there with respect to $s$, giving
$$\sum_{k=0}^\infty {{(-1)^{k+1}} \over {k!}} (s-1)^k \int_0^1 \gamma_{k+1}(a)da
=-{1 \over {(1-s)^2}}.  \eqno(B.10)$$
Relabeling variables, putting $k\to n$ and then putting $s=1-n$ gives the result.

{\it Lemma B3}.  We introduce the quantity $C_n(a)=\gamma_n(a)-{1 \over a}\ln^n a$, that for $n \geq 1$ has the integral representation \cite{coffey2009} 
$$C_n(a)=\int_1^\infty P_1(x-a) {{\ln^{n-1} x} \over x^2}(n-\ln x)dx, \eqno(B.11)$$
where again $P_1(x)$ is the first periodic Bernoulli polynomial.  Then from
Lemma 1 we find
$$A_k(q)=-{1 \over k}-k\sum_{n=0}^\infty {{C_{n+1}(q)} \over {n!}}k^n-kq^{k-1}
\ln q.  \eqno(B.12)$$
Using the $1$-periodicity of $P_1(x)$ yields the relation $A_k(0)=A_k(1)$.

{\it Lemma B4}.  This follows directly from the expansion (1.37).

{\it Supplements to Corollary B1}.  The Bernoulli numbers $B_k$ are well known
to vanish for $k \geq 3$ an odd integer.  That this condition holds in (B.6) is evident upon applying the fact of the trivial zeros of the Riemann zeta
function at $z=-2n$, $n\geq 1$.

Upon the use of $a=1$ in (B.11) inserted in (B.7), we easily recover the known result
$\int_1^\infty P_1(x)dx/x^2=1/2-\gamma$.

{\it Lemma B5}.  This follows from the functional equation
$$\gamma_k(q+1)=\gamma_k(a)-{1 \over q}\ln q.  \eqno(B.13)$$

{\it Lemma B6}.  We use Lemma B1 and Proposition 7 of \cite{coffeyjmaa} so that
$$A_{k+1}'(q)=-(k+1)\sum_{n=0}^\infty {{\gamma_{n+1}'(q)} \over {n!}}(k+1)^n$$
$$=(k+1)\sum_{n=0}^\infty {{(k+1)^n} \over {n!}}(n+1)!(-1)^{n+1} \sum_{\ell=n}^
\infty {{(-1)^\ell} \over {\ell!}} {{\ell+1} \choose {n+1}}\gamma_\ell(q)$$
$$=(k+1)\sum_{n=0}^\infty (k+1)^n (-1)^{n+1}\sum_{\ell=n}^\infty {{(-1)^\ell} \over {\ell!}} {\ell \choose n}(\ell+1)\gamma_\ell(q)$$
$$=-(k+1)\sum_{\ell=0}^\infty {{(\ell+1)} \over \ell!}\gamma_\ell(q)k^\ell$$
$$=-(k+1)\sum_{\ell=0}^\infty {1 \over {\ell !}}[k\gamma_{\ell+1}(q)+\gamma_\ell(q)
]k^\ell.  \eqno(B.14)$$
Above, we reordered the double sum and used the binomial theorem.  Lemma B6 then
follows in light of Lemma B4.

Remark.  Various expressions for values of $A_k$ in \cite{moll} together with
Lemma 1 give Corollaries for certain sums of the Stieltjes constants.  For instance from Lemma 3.5 of \cite{moll} we have 
$$A_k(1/2)=(-1)^{k-1}B_k2^{1-k}\ln 2-(1-2^{1-k})k\zeta'(1-k)
=-{1 \over k}-k\sum_{n=0}^\infty {{\gamma_{n+1}(1/2)} \over {n!}}k^n,  \eqno(B.15)$$
and from Lemma 3.6 there we have
$$A_2(q)=(1-\gamma-\ln 2\pi)(q^2-q+1/6)-{1 \over \pi^2}\sum_{n=1}^\infty {{\ln n}
\over n^2}\cos(2\pi n q)
+{1 \over {2\pi}}\sum_{n=1}^\infty {{\sin(2\pi n q)} \over
n^2}$$
$$=-{1 \over 2}-2\sum_{j=0}^\infty {2^j \over {j!}}\gamma_{j+1}(q).  \eqno(B.16)$$
On the right side of the first line of this equation we recognize the
values Cl$_2(2\pi q)$.

From Lemma B1 we also find the generating function
$$\sum_{k=1}^\infty {{A_k(q)} \over {k!}}z^k=\gamma-\mbox{Ei}(z)+\ln z-ke^z
\sum_{j=0}^\infty {{\gamma_{j+1}(q)} \over {j!}}{\cal B}_j(z), \eqno(B.17)$$
where Ei is the exponential integral and ${\cal B}_j$ the Bell polynomial.

From using the relation $\zeta'(0,s)=\ln \Gamma(s)-(1/2)\ln 2\pi$ and various
integral representations, we have
{\newline \bf Corollary B2}.  We have for Re $s>0$
$$-\int_0^1 {{P_1(x)dx} \over {x+s}}=\int_0^\infty \left({1 \over 2}-{1 \over t}
+{1 \over {e^t-1}}\right){e^{-ts} \over t}dt=2\int_0^\infty {{\tan^{-1}(t/s)} \over
{e^{2\pi t}-1}}dt.  \eqno(B.18)$$
Further Corollaries follow by repeated differentiation with respect to $s$ and
the use of other operations.

We conclude by giving an integral representation for $A_k(q)$ and two
summations relations for it.  We have
{\newline \bf Proposition B1}.  For $k<2$ we have
$$A_k(q)=k\left[-{{\ln q} \over {2q^{1-k}}}-{q^k \over k^2}+{q^k \over k}\ln q
-\int_0^\infty {{P_1(x)dx} \over {(x+q)^{2-k}}}+(1-k)\int_0^\infty {{P_1(x)} \over
{(x+q)^{2-k}}}\ln(x+q)dx\right].  \eqno(B.19)$$
Thus we have
{\newline \bf Corollary B3}.  We have
$$A_1(q)=\left(q-{1 \over 2}\right)\ln q-q-\int_0^\infty {{P_1(x)} \over {x+q}}dx.
\eqno(B.20)$$

For the proof of this Proposition, we apply (2.14) and the definition (B.1).
As far as Corollary B3, it is easy to see that we recover the equivalent of
the relation (e.g., \cite{coffeyjmaa}, (A.2) or \cite{edwards}, p. 107)
$$\ln\Gamma(s+1)=\left(s+{1\over 2}\right)\ln s-s+{1 \over 2}\ln 2\pi-\int_0^\infty
{{P_1(x)dx} \over {x+s}}.  \eqno(B.21)$$

We announce the following summation relations.  The proofs will appear 
elsewhere.  
{\newline \bf Proposition B2}.  Let $p \geq 1$ and $q \geq 1$ be integers, $b\geq 0$, and min$(p/q,q/p) >b$.  Then we have
$$\sum_{r=1}^q A_k\left({{pr} \over q}-b\right)=-\ln\left({q \over p}
\right)\sum_{r=1}^q B_k\left({{pr} \over q}-b\right)+\left({q \over p}\right)^{1-k}
\sum_{\ell=0}^{p-1} A_k\left(1+{{(\ell-b)q} \over p}\right).  \eqno(B.22)$$

{\bf Proposition B3}.  Let ${\cal P}$ denote the prime numbers.  Then for
$p \in {\cal P}$ and integer $N \geq 0$ we have
$$(1-p^{k-1})A_k(1)-(-1)^kp^{k-1} (\ln p)B_k=(-1)^k (N+1)\ln p (1-p^{k-1})B_k$$
$$+p^{(N+1)(k-1)}\sum_{\stackrel{1\leq j<p^{N+1}}{(j,p)=1}}A_k\left({j \over p^{N+1}}\right).  \eqno(B.23)$$

It is possible to derive other such summation relations.

\pagebreak

\end{document}